\documentclass[aps,prd,twocolumn,superscriptaddress,showkeys]{revtex4}
\usepackage{amssymb,amsmath,graphicx,enumerate}
\usepackage{subfigure}
\usepackage{dcolumn,booktabs}
\usepackage{longtable}
\usepackage[dvipdfm,colorlinks=true,citecolor=blue,pdfstartview=FitH]{hyperref}

\usepackage{color,framed} 
\definecolor{shadecolor}{rgb}{0,0,1} 

\def\bea{\begin{equation}}
\def\eea{\end{equation}}
\newcommand{\rt}{Regge trajectory}
\newcommand{\rts}{Regge trajectories}

\newcommand{\trs}{trajectories}
\newcommand{\bfr}{{\bf r}}
\newcommand{\bfp}{{\bf p}}
\newcommand{\bfpa}{{|\bf p|}}
\newcommand{\gev}{{\rm GeV}}

\newcommand{\cltb}{$\bar{3}_c$}
\newcommand{\cltba}{\bar{3}_c}

\newcommand{\dqs}{$(qq')$}

\newcommand{\qp}{{q^{\prime}}}

\newcommand{\qpp}{{\bar{q}^{\prime\prime}}}

\newcommand{\qqs}{{[qq^{\prime}]}}
\newcommand{\qqb}{\{qq^{\prime}\}}

\newcommand{\thts}{triply heavy triquarks}
\newcommand{\tht}{triply heavy triquark}

\begin{document}
\title{Regge trajectories for the triply heavy triquarks}
\author{He Song}
\email{songhe\_22@163.com}
\affiliation{School of Physics and Information Engineering, Shanxi Normal University, Taiyuan 030031, China}
\author{Jia-Qi Xie}
\email{1462718751@qq.com}
\affiliation{School of Physics and Information Engineering, Shanxi Normal University, Taiyuan 030031, China}
\author{Jiao-Kai Chen}
\email{chenjk@sxnu.edu.cn, chenjkphy@outlook.com (corresponding author)}
\affiliation{School of Physics and Information Engineering, Shanxi Normal University, Taiyuan 030031, China}
\begin{abstract}
We attempt to apply the Regge trajectory approach to the triply heavy triquarks $((QQ')\bar{Q}^{\prime\prime})$ $(Q,\,Q',\,Q^{\prime\prime}=b,\,c)$. We present the triquark Regge trajectory relations, and then employ them to crudely estimate the spectra of the triquarks $((cc)\bar{c})$, $((cc)\bar{b})$, $((bc)\bar{c})$, $((bc)\bar{b})$, $((bb)\bar{c})$, and $((bb)\bar{b})$. The $\lambda$-trajectories and the $\rho$-trajectories are discussed.
The triquark Regge trajectory becomes a new and very simple approach for estimating the spectra of triquarks.
Moreover, the spin-averaged masses of the ground states of pentaquarks $(\bar{c}(cc))(cc)$, $(\bar{b}(cc))(cc)$ and $(\bar{c}(bb))(cc)$ are estimated, which are consistent with other theoretical predictions.
\end{abstract}

\maketitle


\section{Introduction}

Triquark correlations are important in spectroscopy and for understanding hadron structure \cite{Karliner:2003dt,Karliner:2003sy,Majee:2007gi,Lee:2004dp,Lebed:2015tna,Giron:2021sla,
Zhu:2015bba,Hogaasen:2004pm,JimenezDelgado:2004rd,Liu:2019zoy,Jaffe:2005md,Yasui:2007dv,
Li:2005rb,Kvinikhidze:2023gmc,Hogaasen:2004ij,Andrew:2023aes}. Triquark can be used to discuss pentaquark and hexaquark \cite{Karliner:2003dt,Karliner:2003sy,Majee:2007gi,Lee:2004dp,Lebed:2015tna,Giron:2021sla,
Zhu:2015bba,Hogaasen:2004pm,JimenezDelgado:2004rd,Andrew:2023aes,Liu:2019zoy}.
Triquarks considered in this work are composed of two quarks and one antiquark. Same as diquarks, triquarks are colored states\cite{Jaffe:2005md}. Although triquarks are not physical,
the triquark spectra have been studied by various methods. In Ref. \cite{Karliner:2003dt}, the triquark $(ud\bar{s})$ mass is obtained by fitting the low-lying mass spectrum.
In Ref. \cite{Lebed:2015tna,Lee:2004dp}, the triquark $((ud)\bar{c})$ mass is estimated by sum rule.
In Ref. \cite{Giron:2021sla}, the triquark $((ud)\bar{c})$ mass is fitted by using the Schr\"{o}dinger equation for different Born-Oppenheimer potentials.
In Ref. \cite{Zhu:2015bba}, the triquark $(ud\bar{c})$ mass is approximated by the sum of three quarks' masses.
In Ref. \cite{Hogaasen:2004pm}, the masses of the triquarks $(ud\bar{s})$ and $(ds\bar{u})$ are given by using the color magnetic Hamiltonian. In Ref. \cite{Yasui:2007dv}, the light triquarks are calculated by the quark model.

Besides these methods, the {\rt} is one of the effective approaches for studying hadron spectra \cite{Regge:1959mz,Chew:1962eu,Nambu:1978bd,Ademollo:1969nx,Baker:2002km,
Brodsky:2006uq,Forkel:2007cm,Filipponi:1997vf,Anisovich:2000kxa,brau:04bs,
Chen:2023web,Chen:2023djq,Chen:2018nnr,Brisudova:1999ut,Masjuan:2012gc,Chen:2014nyo,Guo:2008he,Ebert:2009ub,
Lovelace:1969se,Irving:1977ea,Collins:1971ff,Inopin:1999nf,Afonin:2014nya,Badalian:2019dny,
MartinContreras:2020cyg,Sonnenschein:2018fph,MartinContreras:2023oqs,Roper:2024ovj}. 
In Refs. \cite{Chen:2023ngj,Chen:2023cws,Feng:2023txx}, we apply the {\rt} approach to estimating masses of the colored diquarks. Similar to the diquark case, we attempt to apply the {\rt} approach to crudely estimate the triquark spectra in present work. Our focus here is on the triply heavy triquarks $((cc)\bar{c})$, $((cc)\bar{b})$, $((bc)\bar{c})$, $((bc)\bar{b})$, $((bb)\bar{c})$, and $((bb)\bar{b})$.
To our knowledge, there has not yet been theoretical studies addressing the spectra of the {\thts}. The data obtained from other approaches are expected to check our results.

The paper is organized as follows: In Sec. \ref{sec:rgr}, the {\rt} relations for the {\thts} are obtained from the quadratic form of the spinless Salpeter-type equation. In Sec. \ref{sec:rtdiquark}, we investigate the {\rts} for the {\thts}. The conclusions are presented in Sec. \ref{sec:conc}.

\section{{\rt} relations}\label{sec:rgr}
In this section, the $\lambda$-trajectory and $\rho$-trajectory relations for the {\thts} $((QQ')\bar{Q}^{\prime\prime})$ $(Q,\,Q',\,Q^{\prime\prime}=b,\,c)$ are derived.

\subsection{Preliminary}\label{subsec:prelim}
\begin{figure}[!phtb]
\centering
\includegraphics[width=0.25\textheight]{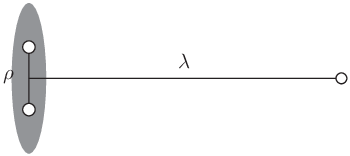}
\caption{Schematic diagram of the triquarks in the antiquark-diquark picture. The grey part represents a diquark composed of two quarks. The circle on the right denotes an antiquark.}\label{fig:tr}
\end{figure}

In the diquark picture, a triquark $((qq^{\prime})\bar{q}^{\prime\prime})$ is regarded as a bound state consisting of one diquark $(qq^{\prime})$ and one antiquark $\bar{q}^{\prime\prime}$ \cite{Lebed:2015tna}, see Fig. \ref{fig:tr}.
The diquark is composed of two quarks. It is in the color antitriplets or sextets, $3_c\otimes3_c=\cltba\oplus{6}_c$. In $SU_c(3)$, there is attraction between quark pairs $(qq')$ in the color antitriplet channel, and this is just twice weaker than in the color singlet $q\bar{q}'$ in the one-gluon exchange approximation \cite{Esposito:2016noz}. Only the color antitriplet states of diquarks are considered here.
One diquark in color {\cltb} and one antiquark in color {\cltb} form a triquark, which will be in a triplet or antisextet in the decomposition of $\cltba\otimes\cltba=3_c\oplus\bar{6}_c$. Only the color triplet states of triquarks are considered \cite{Jaffe:2005md}.

In the diquark picture, a triquark has structure and substructure: $\lambda$ separates the antiquark and the diquark while $\rho$ separates two quarks in the diquark, see Fig. \ref{fig:tr}. There exist two excited modes: the $\rho$-mode involves the radial and orbital excitation in the diquark, and the $\lambda$-mode involves the radial or orbital excitation between the antiquark and diquark. Consequently, there exist two series of {\rts}: one series of $\rho$-{\trs} and one series of $\lambda$-{\trs}.

\begin{table}[!phtb]
\caption{The completely antisymmetric states for the diquarks in {\cltb} \cite{Feng:2023txx}. $j_d$ is the spin of the diquark {\dqs}, $s_d$ denotes the total spin of two quarks, $l$ represents the orbital angular momentum. $n=n_r+1$, $n_r$ is the radial quantum number, $n_r=0,1,2,\cdots$. $q$ and $q'$ denote both the light quarks and the heavy quarks.}  \label{tab:dqstates}
\centering
\begin{tabular*}{0.49\textwidth}{@{\extracolsep{\fill}}ccccc@{}}
\hline\hline
 Spin of diquark & Parity  &  Wave state  &  Configuration    \\
( $j_d$ )          & $(j_d^P)$ & $(n^{2s_d+1}l_{j_d})$  \\
\hline
$j_d=0$              & $0^+$   & $n^1s_0$         & $[qq']^{{\cltba}}_{n^1s_0}$ \\
                 & $0^-$   & $n^3p_0$         & $[qq']^{{\cltba}}_{n^3p_0}$       \\
$j_d=1$              & $1^+$   & $n^3s_1$, $n^3d_1$   & $\{qq'\}^{{\cltba}}_{n^3s_1}$,\;    $\{qq'\}^{{\cltba}}_{n^3d_1}$\\
                 & $1^-$   & $n^1p_1$, $n^3p_1$   &
$\{qq'\}^{{\cltba}}_{n^1p_1}$,\; $[qq']^{{\cltba}}_{n^3p_1}$ \\
$j_d=2$              & $2^+$   & $n^1d_2$, $n^3d_2$         &  $[qq']^{{\cltba}}_{n^1d_2}$,\; $\{qq'\}^{{\cltba}}_{n^3d_2}$\\
                 & $2^-$   & $n^3p_2$, $n^3f_2$       &
 $[qq']^{{\cltba}}_{n^3p_2}$,\; $[qq']^{{\cltba}}_{n^3f_2}$       \\
$\cdots$         & $\cdots$ & $\cdots$               & $\cdots$  \\
\hline\hline
\end{tabular*}
\end{table}

In the diquark picture, the state of a triquark is denoted as
\bea\label{tetnot}
\left((qq')^{\cltba}_{n^{2s_d+1}l_{j_d}}\bar{q}^{\prime\prime}\right)^{3_c}_{N^{2j+1}L_J}.
\eea
The diquark $(qq')$ is $\{qq'\}$ or $[qq']$. $\{qq'\}$ and $[qq']$ denote the symmetric and antisymmetric flavor wave functions, respectively. The completely antisymmetric states for the diquarks in {\cltb} are listed in Table \ref{tab:dqstates}. $N=N_{r}+1$, where $N_{r}=0,\,1,\,\cdots$. $n=n_{r}+1$, where $n_{r}=0,\,1,\,\cdots$. $N_r$ and $n_{r}$ are the radial quantum numbers of the triquark and diquark, respectively.
$\vec{J}=\vec{L}+\vec{j}$, $\vec{j}=\vec{j}_d+\vec{s}_{\qpp}$, $\vec{j}_d=\vec{s}_d+\vec{l}$.
$\vec{J}$, $\vec{j}_d$ and $\vec{s}_{\qpp}$ are the spins of triquark, diquark and antiquark $q^{\prime\prime}$, respectively. $\vec{j}$ is the summed spin of diquark and antiquark in the triquark. $L$ and $l$ are the orbital quantum numbers of triquark and diquark, respectively. $\vec{s}_{d}$ is the summed spin of quarks in the diquark.

\subsection{QSSE and the {\rt} relation}

In Ref. \cite{Feng:2023txx}, we show that the {\rts} for the doubly heavy diquarks can be described by the ansatz \cite{Chen:2022flh}
\bea\label{massf}
M=\beta_x(x+c_{0x})^{\nu}+m_R,\,(x=l,\,n_r).
\eea
In this subsection, we will show that the {\rts} for the triply heavy triquarks can also be described by this ansatz.

The quadratic form of the spinless Salpeter-type equation reads as \cite{Baldicchi:2007ic,Baldicchi:2007zn,Brambilla:1995bm,
chenvp,chenrm,Chen:2018hnx,Chen:2018bbr,Chen:2018nnr,Chen:2021kfw}
\begin{eqnarray}\label{qsse}
M^2\Psi_{d,t}({\bfr})=M_0^2\Psi({\bfr})+{\mathcal U}_{d,t}\Psi_{d,t}({\bfr}),\quad M_0=\omega_1+\omega_2,
\end{eqnarray}
where $\Psi_{d,t}$ are the diquark wave function and the triquark wave function, respectively. $\omega_1$ is the relativistic energy of constituent $1$ (quark $q$ or diquark $(qq')$), and $\omega_2$ is of constituent $2$ (quark $q'$ or antiquark $\bar{q}^{\prime\prime}$),
\bea\label{omega}
\omega_i=\sqrt{m_i^2+{\bf p}^2}=\sqrt{m_i^2-\Delta},
\eea
\bea\label{potu}
{\mathcal U}=M_0V_{d,t}+V_{d,t}M_0+V_{d,t}^2.
\eea
$m_1$ and $m_2$ are the effective masses of constituent $1$ and constituent $2$, respectively.

Following Refs. \cite{Ferretti:2019zyh,Bedolla:2019zwg,Ferretti:2011zz,Eichten:1974af,
Chen:2023web,Chen:2023djq}, we employ the potential
\begin{align}\label{potv}
V_{d,t}&=-\frac{3}{4}\left[V_c+{\sigma}r+C\right]
\left({\bf{F}_i}\cdot{\bf{F}_j}\right)_{d,t},
\end{align}
where $V_c\propto{1/r}$ is a color Coulomb potential or a Coulomb-like potential due to one-gluon-exchange. $\sigma$ is the string tension. $C$ is a fundamental parameter \cite{Gromes:1981cb,Lucha:1991vn}. The part in the bracket is the Cornell potential \cite{Eichten:1974af}. ${\bf{F}_i}\cdot{\bf{F}_j}$ is the color-Casimir,
\bea\label{mrcc}
\langle{(\bf{F}_i}\cdot{\bf{F}_j})_{d,t}\rangle=-\frac{2}{3}.
\eea

For the heavy-heavy systems, $m_{1},m_2{\gg}{\bfpa}$, Eq. (\ref{qsse}) reduces to
\begin{eqnarray}\label{qssenr}
M^2\Psi_{d,t}({\bfr})&=&\left[(m_1+m_2)^2+\frac{m_1+m_2}{\mu}{\bfp}^2\right]\Psi_{d,t}({\bfr})\nonumber\\
&&+2(m_1+m_2)V_{d,t}\Psi_{d,t}({\bfr})+\cdots,
\end{eqnarray}
where
\bea
\mu=m_1m_2/(m_1+m_2).
\eea
The $\cdots$ in Eq. (\ref{qssenr}) represents $V_{d,t}^2$, which is at the same order as the term $(m_1+m_2){\bfp}^2/\mu$. When determining the form of the {\rt}, $V_{d,t}^2$ is omitted because it is a small term compared to $2(m_1+m_2)V_{d,t}$. The form of the {\rt} is determined by the terms $(m_1+m_2){\bfp}^2/\mu$ and $2(m_1+m_2)V_{d,t}$ while $V_{d,t}^2$ provides a correction that we do not consider here. The {\rts} (\ref{massf}) and (\ref{massform}) obtained from Eq. (\ref{qssenr}) are consistent with the results obtained from the spinless Salpeter equation and other models; see Refs. \cite{Chen:2022flh,Chen:2021kfw} for more details.
By employing the Bohr-Sommerfeld quantization approach \cite{Brau:2000st} and using Eqs. (\ref{potv}) and (\ref{qssenr}), we can obtain (\ref{massf}) with the following parameters \cite{Feng:2023txx,Chen:2022flh,Chen:2018hnx,Chen:2021kfw}
\begin{align}\label{massform}
\nu&=2/3,\quad \beta_x=c_{fx}c_xc_c,\quad  x=l,\,n_r,\,L,\,N_r,\nonumber\\
m_R&=m_1+m_2+C'.
\end{align}
The constants $c_{x}$ and $c_c$ are
\begin{align}\label{cxcons}
c_c&=\left(\frac{\sigma'^2}{\mu}\right)^{1/3},\quad c_{l,L}=\frac{3}{2},\quad c_{n_r,N_r}=\frac{\left(3\pi\right)^{2/3}}{2},\nonumber\\
C'&=\frac{C}{2},\quad \sigma'=\frac{\sigma}{2}.
\end{align}
$c_{fx}$ are theoretically equal to 1 and are fitted in practice.
In Eqs. (\ref{massform}) and (\ref{cxcons}), $m_1$, $m_2$, $C$, $c_x$, $c_{fx}$ and $\sigma$ are universal for the doubly heavy diquarks and the triply heavy triquarks. $c_{0x}$ is determined by fitting a given {\rt}.
If the confining potential is not linear, the exponent $\nu$ will change \cite{Feng:2023txx}.

\subsection{{\rt} relations for the {\thts}  }

A {\tht} consists of one doubly heavy diquark and one heavy antiquark, therefore, it is a heavy-heavy system for the $\lambda$-mode. One of the constituents of the {\tht}, the doubly heavy diquark composed of two heavy quark, is also a heavy-heavy system. Using formulas (\ref{massf}), (\ref{massform}) and (\ref{cxcons}), we have the {\rt} relations for the {\thts}
\begin{align}\label{t2q}
M&=m_{R_{\lambda}}+\beta_{x_{\lambda}}(x_{\lambda}+c_{0x_{\lambda}})^{2/3}\;(x_{\lambda}=L,\,N_r),\nonumber\\
M_{d}&=m_{R_\rho}+\beta_{x_{\rho}}(x_{\rho}+c_{0x_{\rho}})^{2/3}\;(x_{\rho}=l,\,n_{r}),
\end{align}
where
\begin{align}\label{pa2qQ}
m_{R_{\lambda}}&=M_{d}+m_{q^{\prime\prime}}+C/2,\nonumber\\
m_{R_\rho}&=m_{q}+m_{q'}+C/2,\nonumber\\
\beta_{L}&=\frac{3}{2}\left(\frac{\sigma^2}{4\mu_{\lambda}}\right)^{1/3}c_{fL},\; \beta_{N_r}=\frac{(3\pi)^{2/3}}{2}\left(\frac{\sigma^2}{4\mu_{\lambda}}\right)^{1/3}c_{fN_r},\nonumber\\
\mu_{\lambda}&=\frac{M_{d}m_{q^{\prime\prime}}}{M_{d}+m_{q^{\prime\prime}}},\;
\mu_{\rho}=\frac{m_{q}m_{q^{\prime}}}{m_{q}+m_{q^{\prime}}},\nonumber\\
\beta_{l}&=\frac{3}{2}\left(\frac{\sigma^2}{4\mu_{\rho}}\right)^{1/3}c_{fl},\; \beta_{n_r}=\frac{(3\pi)^{2/3}}{2}\left(\frac{\sigma^2}{4\mu_{\rho}}\right)^{1/3}c_{fn_r}.
\end{align}
In Eq. (\ref{t2q}), $M$ is the triquark mass, and $M_{d}$ is the diquark mass. The second relation in Eq. (\ref{t2q}) is used to calculate the diquark masses \cite{Feng:2023txx}. The relations in Eqs. (\ref{t2q}) and (\ref{pa2qQ}) are employed to calculate the triquark mass.

According to Eqs. (\ref{t2q}) and (\ref{pa2qQ}), we have
\bea
M=M_{d}+m_{q^{\prime\prime}}+C/2+\beta_{x_{\lambda}}(x_{\lambda}+c_{0x_{\lambda}})^{2/3}
\eea
when the diquark is regarded as a constituent and the structure of the diquark is not considered, then we have the binding energies of the {\thts}, $\epsilon=C/2+\beta_{x_{\lambda}}(x_{\lambda}+c_{0x_{\lambda}})^{2/3}$. When the diquark is considered as a bound state composed of two heavy quarks \cite{Feng:2023txx}, we have
\begin{align}\label{combrt}
M=&m_{q}+m_{q'}+m_{q^{\prime\prime}}+C\nonumber\\
&+\beta_{x_{\lambda}}(x_{\lambda}+c_{0x_{\lambda}})^{2/3}
+\beta_{x_{\rho}}(x_{\rho}+c_{0x_{\rho}})^{2/3}
\end{align}
from Eqs. (\ref{t2q}) and (\ref{pa2qQ}), then the binding energies of the {\thts} read $\epsilon=C+\beta_{x_{\lambda}}(x_{\lambda}+c_{0x_{\lambda}})^{2/3}+
\beta_{x_{\rho}}(x_{\rho}+c_{0x_{\rho}})^{2/3}$.
We can see from Eq. (\ref{combrt}) that there are two series of {\rts} for the {\thts}: the $\lambda$-trajectories as $x_{\rho}$ is fixed and the $\rho$-trajectories as $x_{\lambda}$ is fixed.
The {\rt} relations [Eqs. (\ref{t2q}) and (\ref{pa2qQ})] for the {\thts} have the same form as the {\rt} relations for the triply heavy baryons \cite{Xie:2024lfo}.

\section{{\rts} for the {\thts}}\label{sec:rtdiquark}

In this section, the {\rts} for the {\thts} $((cc)\bar{c})$, $((cc)\bar{b})$, $((bc)\bar{c})$, $((bc)\bar{b})$, $((bb)\bar{c})$, and $((bb)\bar{b})$ are investigated by using Eqs. (\ref{t2q}) and (\ref{pa2qQ}) or Eqs. (\ref{combrt}) and (\ref{pa2qQ}).

\subsection{Parameters}

\begin{table}[!phtb]
\caption{The values of parameters \cite{Feng:2023txx,Faustov:2021qqf}.}  \label{tab:parmv}
\centering
\begin{tabular*}{0.45\textwidth}{@{\extracolsep{\fill}}cc@{}}
\hline\hline
          & $m_{c}=1.55\; {\gev}$, \; $m_b=4.88\; {\gev}$, \\
          & $\sigma=0.18\; {\gev^2}$,\; $C=-0.3\; {\gev}$, \\
          & $c_{fn_{r}}=1.0$,\; $c_{fl}=1.17$        \\
$(cc)$    & $c_{0n_{r}}(1^3s_1)=0.205$,\quad $c_{0{l}}(1^3s_1)=0.337$,\\
$(bc)$    & $c_{0n_{r}}(1^3s_1)=0.182$,\quad $c_{0{l}}(1^3s_1)=0.257$,\\
          & $c_{0n_{r}}(1^1s_0)=0.107$,\quad $c_{0{l}}(1^1s_0)=0.169$,\\
$(bb)$    & $c_{0n_{r}}(1^3s_1)=0.01$,\quad  $c_{0{l}}(1^3s_1)=0.001$.\\
\hline
\hline
\end{tabular*}
\end{table}

The parameter values are listed in Table \ref{tab:parmv}. The values of $m_b$, $m_c$, $\sigma$ and $C$ are taken directly from \cite{Faustov:2021qqf}. $c_{fx}$ and $c_{0x}$ for the $\rho$-mode are obtained by fitting the {\rts} for the doubly heavy mesons, and then are used to fit the {\rts} for the doubly heavy diquarks. $c_{fx}$ are universal for all doubly heavy diquark {\rts} while $c_{0x}$ varies with different diquark {\rts} \cite{Feng:2023txx}.
The parameters for the $\lambda$-mode are determined by the relations \cite{Xie:2024dfe}
\begin{eqnarray}
c_{fL}=&1.116 + 0.013\mu_{\lambda},\; c_{0L}=0.540- 0.141\mu_{\lambda}, \nonumber\\
c_{fN_r}=&1.008 + 0.008\mu_{\lambda}, \;  c_{0N_r}=0.334 - 0.087\mu_{\lambda},\label{fitcfxnr}
\end{eqnarray}
where $\mu_{\lambda}$ is the reduced masses, see Eq. (\ref{pa2qQ}). The relations in Eq. (\ref{fitcfxnr}) are obtained by fitting the mesons, baryons and tetraquarks. Because the triquarks are not physical, there are not experimental data for the triquark masses. Therefore, these parameters in (\ref{fitcfxnr}) cannot be determined by using triquark masses. We use the relations as a provisional method before finding a better one. It can be validated by whether the fitted results for the triquarks agree with the theoretical values obtained by using other approaches.

\subsection{{\rts} for the triquarks $((cc)\bar{c})$ and $((cc)\bar{b})$}\label{subsec:rtsa}

\begin{table}[!phtb]
\caption{The spin-averaged masses of the $\lambda$-excited states of $((cc)\bar{c})$ and $((cc)\bar{b})$ (in ${\gev}$). The notation in Eq. (\ref{tetnot}) is rewritten as $|n^{2s_d+1}l_{j_d},N^{2j+1}L_J\rangle$. And $|n^{2s_d+1}l_{j_d},NL\rangle$ denotes the spin-averaged states. Eqs. (\ref{t2q}), (\ref{pa2qQ}) and (\ref{fitcfxnr}) are used.}  \label{tab:masslambda}
\centering
\begin{tabular*}{0.50\textwidth}{@{\extracolsep{\fill}}ccc@{}}
\hline\hline
\addlinespace[2pt]
  $|n^{2s_d+1}l_{j_d},NL\rangle$ & $((cc)\bar{c})$  & $((cc)\bar{b})$ \\
\hline
 $|1^3s_1, 1S\rangle$  &4.70    &7.96      \\
 $|1^3s_1, 2S\rangle$  &5.04    &8.26   \\
 $|1^3s_1, 3S\rangle$  &5.29    &8.47   \\
 $|1^3s_1, 4S\rangle$  &5.51    &8.65   \\
 $|1^3s_1, 5S\rangle$  &5.70    &8.81   \\
\hline
 $|1^3s_1, 1S\rangle$  &4.70    &7.97   \\
 $|1^3s_1, 1P\rangle$  &4.94    &8.18   \\
 $|1^3s_1, 1D\rangle$  &5.12    &8.33  \\
 $|1^3s_1, 1F\rangle$  &5.28    &8.46   \\
 $|1^3s_1, 1G\rangle$  &5.42    &8.58   \\
 $|1^3s_1, 1H\rangle$  &5.55    &8.69  \\
\hline\hline
\end{tabular*}
\end{table}

\begin{table}[!htbp]
\caption{Same as Table \ref{tab:masslambda} except for the $\rho$-excited states. $\times$ denotes the nonexist states.} \label{tab:massrho}
\centering
\begin{tabular*}{0.5\textwidth}{@{\extracolsep{\fill}}ccc@{}}
\hline\hline
\addlinespace[2pt]
  $|n^{2s_d+1}l_{j_d},NL\rangle$ & $((cc)\bar{c})$  & $((cc)\bar{b})$ \\
\hline
 $|1^3s_1, 1S\rangle$  &4.70    &7.96   \\
 $|2^3s_1, 1S\rangle$  &5.07    &8.34   \\
 $|3^3s_1, 1S\rangle$  &5.35    &8.61   \\
 $|4^3s_1, 1S\rangle$  &5.58    &8.84   \\
 $|5^3s_1, 1S\rangle$  &5.79    &9.04   \\
\hline
 $|1^3s_1, 1S\rangle$  &4.71    &7.98   \\
 $|1^3p_2, 1S\rangle$$(\times)$  &4.99    &8.25   \\
 $|1^3d_3, 1S\rangle$  &5.20    &8.46  \\
 $|1^3f_4, 1S\rangle$$(\times)$  &5.38    &8.64   \\
 $|1^3g_5, 1S\rangle$  &5.54    &8.80  \\
 $|1^3h_6, 1S\rangle$$(\times)$ &5.69    &8.95  \\
\hline\hline
\end{tabular*}
\end{table}

\begin{figure*}[!phtb]
\centering
\subfigure[]{\label{subfigure:cfa}\includegraphics[scale=0.48]{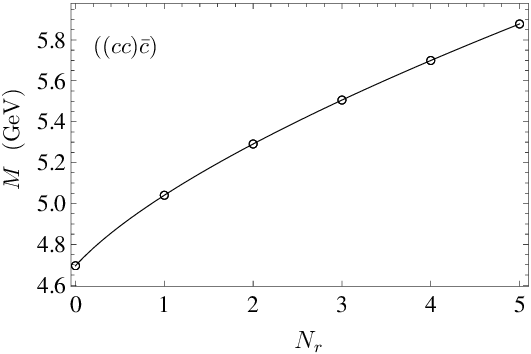}}
\subfigure[]{\label{subfigure:cfa}\includegraphics[scale=0.48]{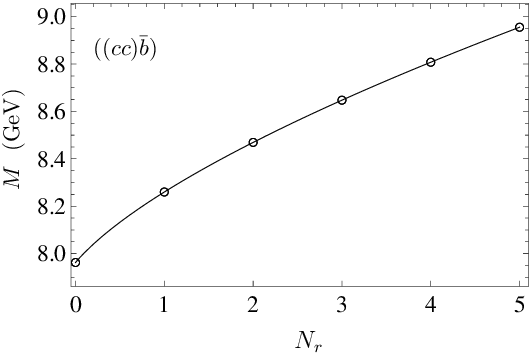}}
\subfigure[]{\label{subfigure:cfa}\includegraphics[scale=0.48]{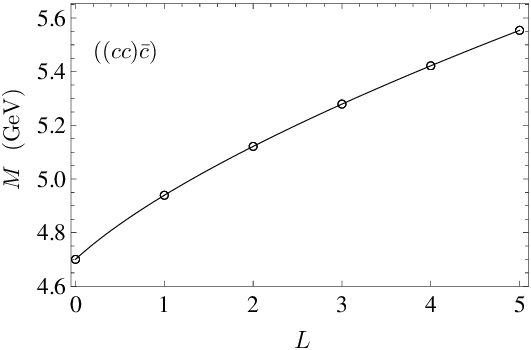}}
\subfigure[]{\label{subfigure:cfa}\includegraphics[scale=0.48]{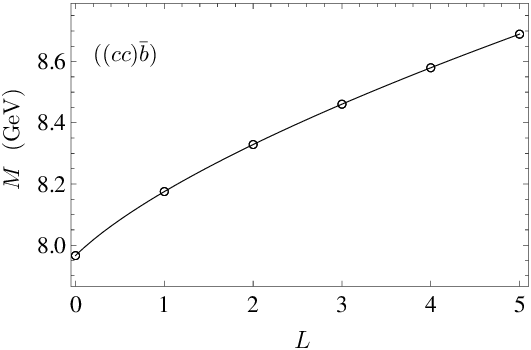}}
\subfigure[]{\label{subfigure:cfa}\includegraphics[scale=0.48]{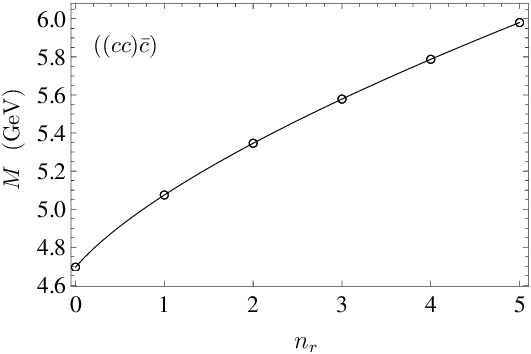}}
\subfigure[]{\label{subfigure:cfa}\includegraphics[scale=0.48]{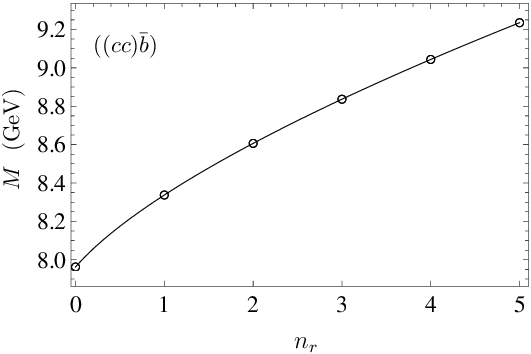}}
\subfigure[]{\label{subfigure:cfa}\includegraphics[scale=0.48]{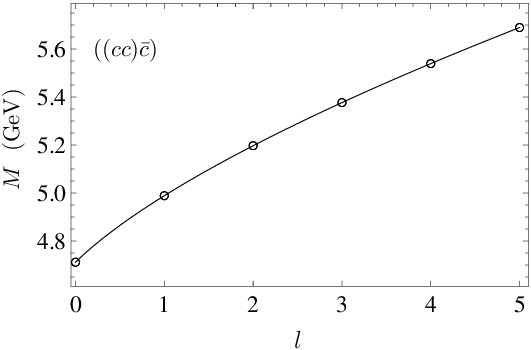}}
\subfigure[]{\label{subfigure:cfa}\includegraphics[scale=0.48]{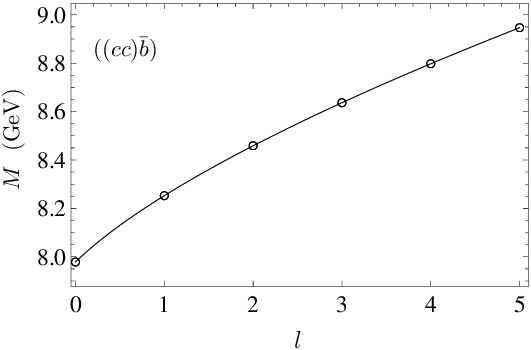}}
\caption{The $\lambda$- and $\rho$-trajectories for the triquarks $((cc)\bar{c})$ and $((cc)\bar{b})$. The radial and orbital $\lambda$-trajectories are plotted in the $(M,\,N_r)$ plane and in the $(M,\,L)$ plane, respectively. The radial and orbital $\rho$-trajectories are plotted in the $(M,\,n_r)$ plane and in the $(M,\,l)$ plane, respectively. Circles represent the predicted data and the black lines are the {\rts}, see Eq. (\ref{t2q}). Data are listed in Tables \ref{tab:masslambda} and \ref{tab:massrho}.}\label{fig:cc}
\end{figure*}

\begin{figure*}[!phtb]
\centering
\subfigure[]{\label{subfigure:cfa}\includegraphics[scale=0.48]{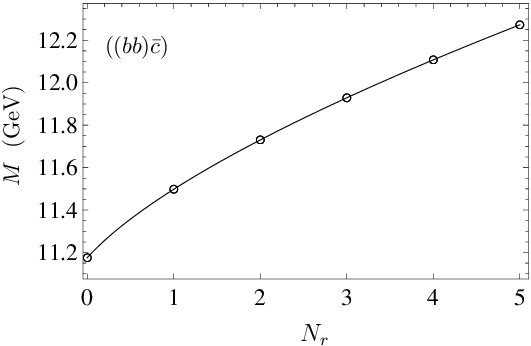}}
\subfigure[]{\label{subfigure:cfa}\includegraphics[scale=0.48]{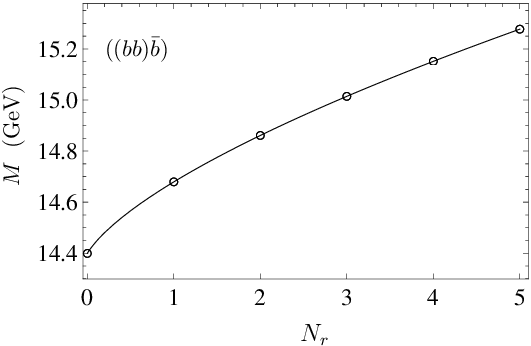}}
\subfigure[]{\label{subfigure:cfa}\includegraphics[scale=0.48]{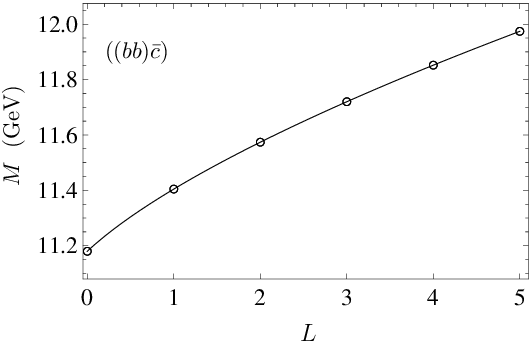}}
\subfigure[]{\label{subfigure:cfa}\includegraphics[scale=0.48]{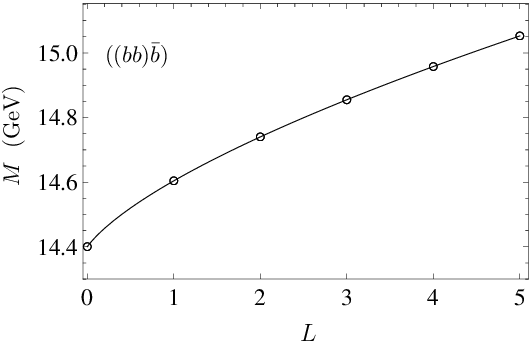}}
\subfigure[]{\label{subfigure:cfa}\includegraphics[scale=0.48]{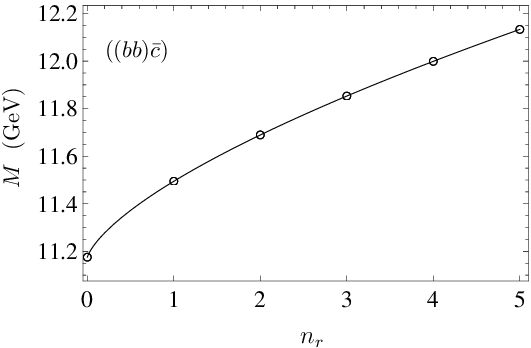}}
\subfigure[]{\label{subfigure:cfa}\includegraphics[scale=0.48]{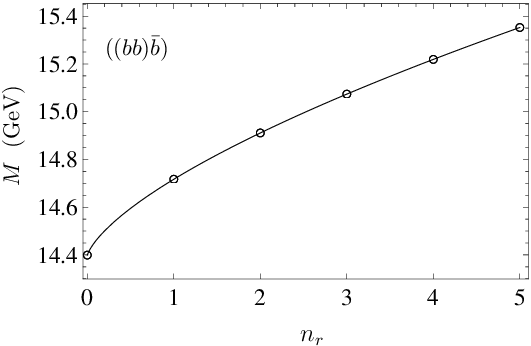}}
\subfigure[]{\label{subfigure:cfa}\includegraphics[scale=0.48]{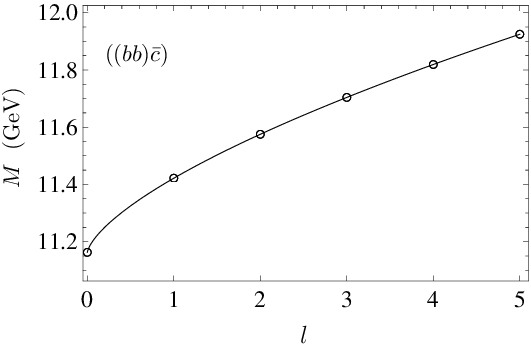}}
\subfigure[]{\label{subfigure:cfa}\includegraphics[scale=0.48]{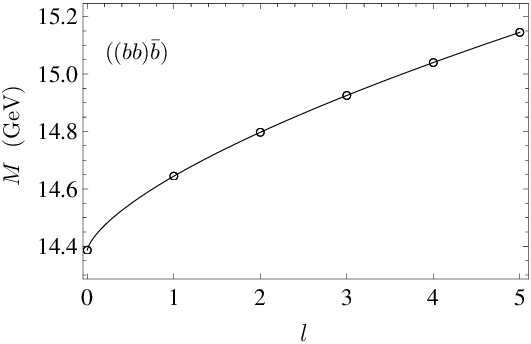}}
\caption{Same as Fig. \ref{fig:cc} except for the triquarks $((bb)\bar{c})$ and $((bb)\bar{b})$, and data are listed in Tables \ref{tab:masslambdab} and \ref{tab:massrhob}.}\label{fig:bb}
\end{figure*}

When calculating the $\lambda$-mode radially and orbitally excited states, the $\rho$-mode state is taken as the radial ground state. Similarly, when calculating the $\rho$-mode radially and orbitally excited states, the $\lambda$-mode state is taken as the radial ground state.

Using Eqs. (\ref{t2q}), (\ref{pa2qQ}), and (\ref{fitcfxnr}), and parameters in Table \ref{tab:parmv}, the spectra of the $\lambda$-excited states and the $\rho$-excited states of the triquarks $((cc)\bar{c})$ and $((cc)\bar{b})$ can be calculated, see Tables \ref{tab:masslambda} and \ref{tab:massrho}. Both the $\lambda$- and $\rho$-trajectories for the triquarks $((cc)\bar{c})$ and $((cc)\bar{b})$ are shown in Fig. \ref{fig:cc}.

\subsection{{\rts} for the triquarks $((bb)\bar{c})$ and $((bb)\bar{b})$}\label{subsec:rtsb}

\begin{table}[!phtb]
\caption{Same as Table \ref{tab:masslambda} except for the triquarks $((bb)\bar{c})$ and $((bb)\bar{b})$.}  \label{tab:masslambdab}
\centering
\begin{tabular*}{0.50\textwidth}{@{\extracolsep{\fill}}ccc@{}}
\hline\hline
\addlinespace[2pt]
  $|n^{2s_d+1}l_{j_d},NL\rangle$ & $((bb)\bar{c})$  & $((bb)\bar{b})$ \\
\hline
 $|1^3s_1, 1S\rangle$  &11.18    &14.40      \\
 $|1^3s_1, 2S\rangle$  &11.50    &14.68   \\
 $|1^3s_1, 3S\rangle$  &11.73    &14.86   \\
 $|1^3s_1, 4S\rangle$  &11.93    &15.01   \\
 $|1^3s_1, 5S\rangle$  &12.11    &15.15  \\
\hline
 $|1^3s_1, 1S\rangle$  &11.18    &14.40  \\
 $|1^3s_1, 1P\rangle$  &11.40    &14.60   \\
 $|1^3s_1, 1D\rangle$  &11.57    &14.74   \\
 $|1^3s_1, 1F\rangle$  &11.72    &14.86   \\
 $|1^3s_1, 1G\rangle$  &11.85    &14.96   \\
 $|1^3s_1, 1H\rangle$  &11.97    &15.05  \\
\hline\hline
\end{tabular*}
\end{table}

\begin{table}[!htbp]
\caption{Same as Table \ref{tab:massrho} except for the triquarks $((bb)\bar{c})$ and $((bb)\bar{b})$.}  \label{tab:massrhob}
\centering
\begin{tabular*}{0.5\textwidth}{@{\extracolsep{\fill}}ccc@{}}
\hline\hline
\addlinespace[2pt]
  $|n^{2s_d+1}l_{j_d},NL\rangle$ & $(bb)\bar{c}$  & $(bb)\bar{b}$ \\
\hline
 $|1^3s_1, 1S\rangle$  &11.18    &14.40   \\
 $|2^3s_1, 1S\rangle$  &11.49    &14.72  \\
 $|3^3s_1, 1S\rangle$  &11.69    &14.91   \\
 $|4^3s_1, 1S\rangle$  &11.85    &15.07  \\
 $|5^3s_1, 1S\rangle$  &12.00    &15.22   \\
\hline
 $|1^3s_1, 1S\rangle$  &11.16    &14.39   \\
 $|1^3p_2, 1S\rangle$$(\times)$  &11.42    &14.64   \\
 $|1^3d_3, 1S\rangle$  &11.58    &14.80  \\
 $|1^3f_4, 1S\rangle$$(\times)$ &11.70    &14.93   \\
 $|1^3g_5, 1S\rangle$  &11.82    &15.04   \\
 $|1^3h_6, 1S\rangle$$(\times)$  &11.92    &15.15  \\
\hline\hline
\end{tabular*}
\end{table}

Similar to the case of triquarks $((cc)\bar{c})$ and $((cc)\bar{b})$, the spectra of and the {\rts} for triquarks $((bb)\bar{c})$ and $((bb)\bar{b})$ can be obtained.
Employing Eqs. (\ref{t2q}), (\ref{pa2qQ}), and (\ref{fitcfxnr}), and parameters in Table \ref{tab:parmv}, the spectra of the $\lambda$-excited states and the $\rho$-excited states of the triquarks $((bb)\bar{c})$ and $((bb)\bar{b})$ are calculated, see Tables \ref{tab:masslambdab} and \ref{tab:massrhob}. The $\lambda$- and $\rho$-trajectories for the triquarks $((bb)\bar{c})$ and $((bb)\bar{b})$ are shown in Fig. \ref{fig:bb}.

\subsection{{\rts} for the triquarks $((bc)\bar{c})$ and $((bc)\bar{b})$}\label{subsec:rtsc}

\begin{figure*}[!phtb]
\centering
\subfigure[]{\label{subfigure:cfa}\includegraphics[scale=0.48]{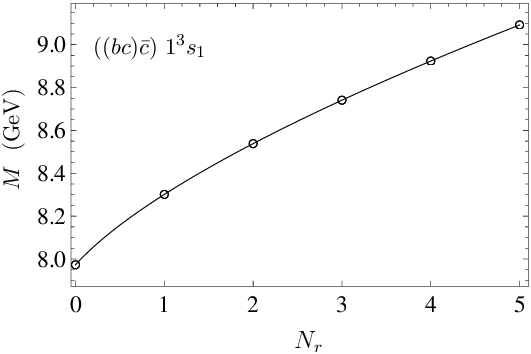}}
\subfigure[]{\label{subfigure:cfa}\includegraphics[scale=0.48]{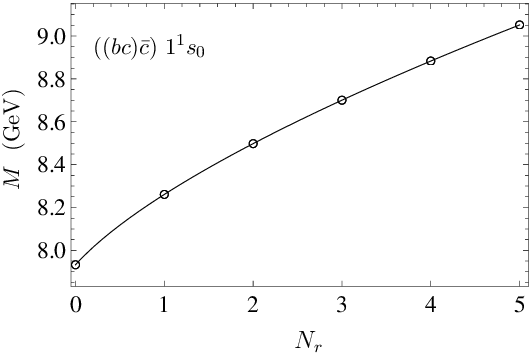}}
\subfigure[]{\label{subfigure:cfa}\includegraphics[scale=0.48]{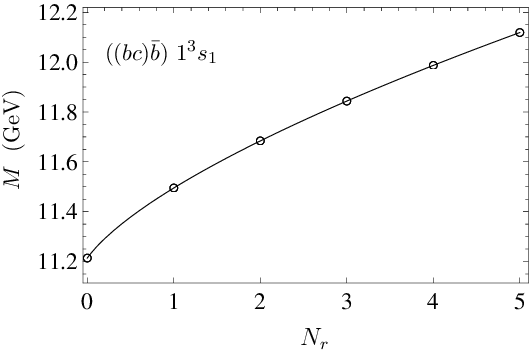}}
\subfigure[]{\label{subfigure:cfa}\includegraphics[scale=0.48]{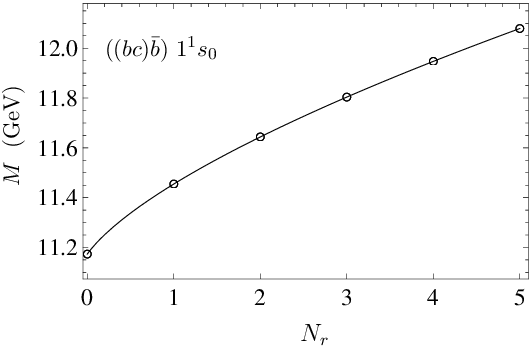}}
\subfigure[]{\label{subfigure:cfa}\includegraphics[scale=0.48]{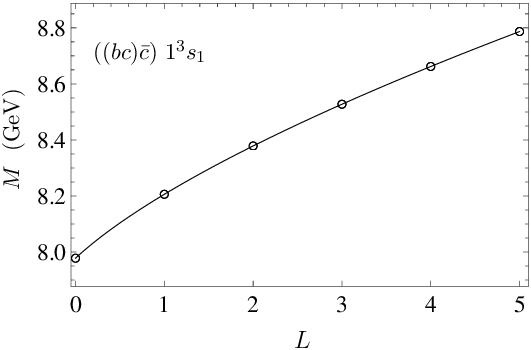}}
\subfigure[]{\label{subfigure:cfa}\includegraphics[scale=0.48]{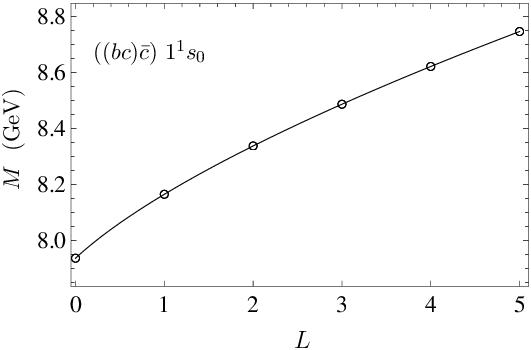}}
\subfigure[]{\label{subfigure:cfa}\includegraphics[scale=0.48]{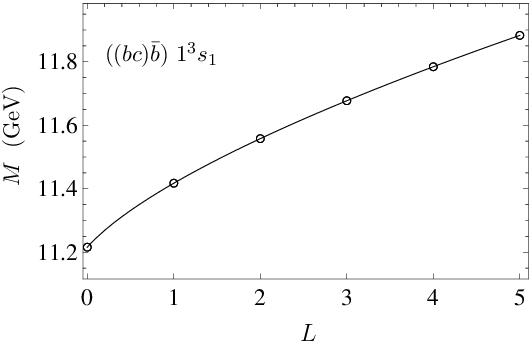}}
\subfigure[]{\label{subfigure:cfa}\includegraphics[scale=0.48]{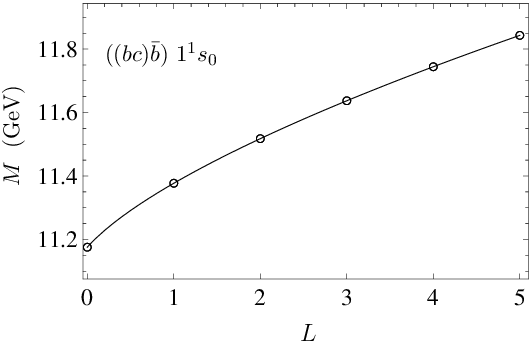}}
\subfigure[]{\label{subfigure:cfa}\includegraphics[scale=0.48]{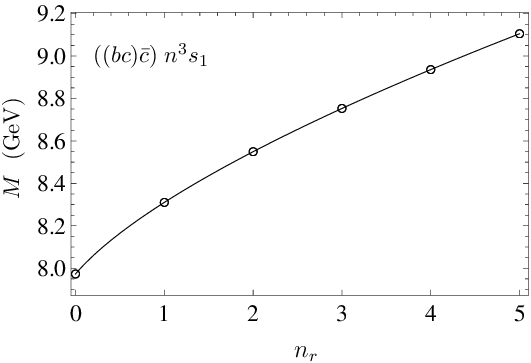}}
\subfigure[]{\label{subfigure:cfa}\includegraphics[scale=0.48]{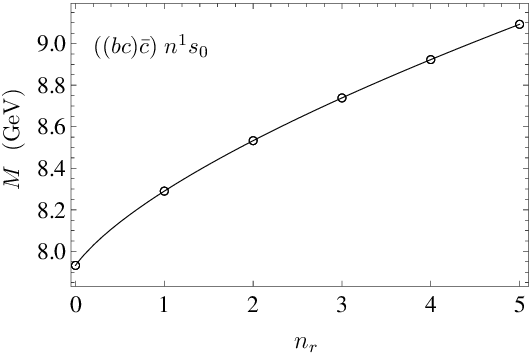}}
\subfigure[]{\label{subfigure:cfa}\includegraphics[scale=0.48]{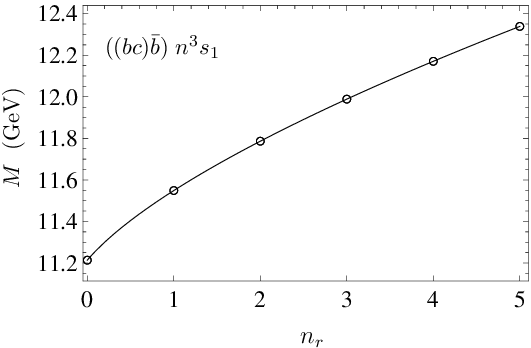}}
\subfigure[]{\label{subfigure:cfa}\includegraphics[scale=0.48]{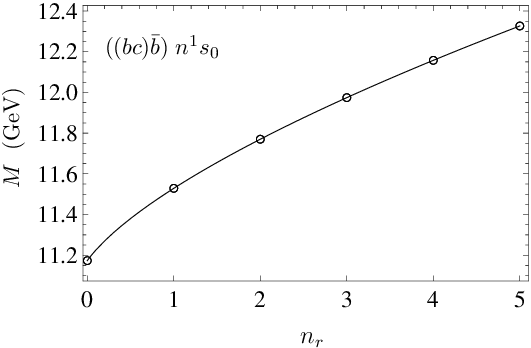}}
\subfigure[]{\label{subfigure:cfa}\includegraphics[scale=0.48]{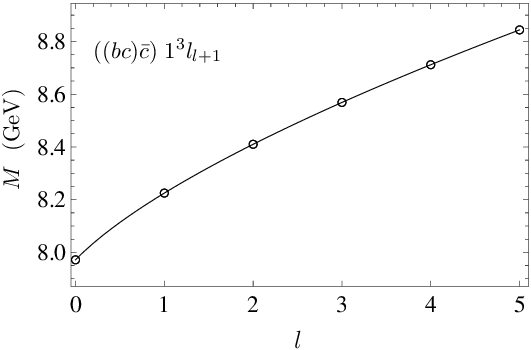}}
\subfigure[]{\label{subfigure:cfa}\includegraphics[scale=0.48]{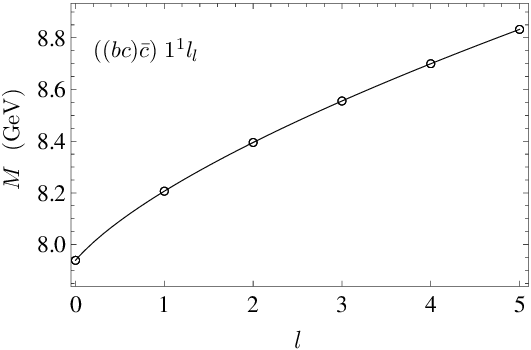}}
\subfigure[]{\label{subfigure:cfa}\includegraphics[scale=0.48]{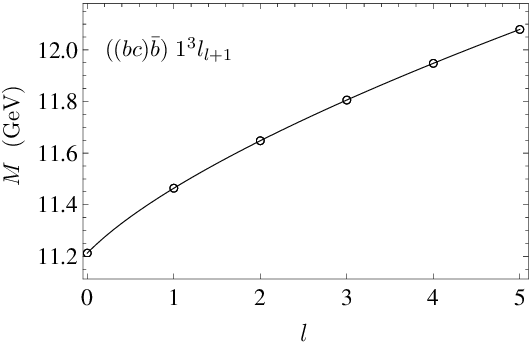}}
\subfigure[]{\label{subfigure:cfa}\includegraphics[scale=0.48]{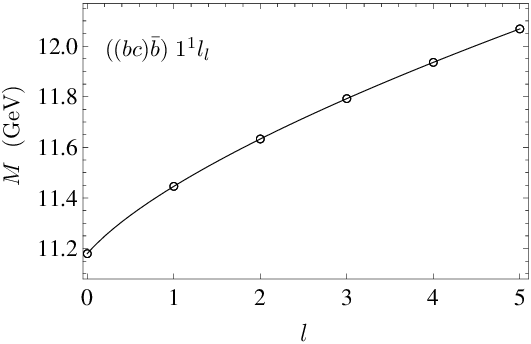}}
\caption{Same as Fig. \ref{fig:cc} except for the triquarks $((bc)\bar{c})$ and $((bc)\bar{b})$, and data are listed in Tables \ref{tab:masslambdac} and \ref{tab:massrhoc}. $1^1s_0$, $n^1s_0$ and $1^1l_{l}$ denote the spin singlets of diquarks. $1^3s_1$, $n^3s_1$ and $1^3l_{l+1}$ denote the spin triplets.}\label{fig:bc}
\end{figure*}

Employing Eqs. (\ref{t2q}), (\ref{pa2qQ}), and (\ref{fitcfxnr}), and parameters in Table \ref{tab:parmv}, the spectra of the $\lambda$-excited states and the $\rho$-excited states of the triquarks $((bc)\bar{c})$ and $((bc)\bar{b})$ are calculated, see Tables \ref{tab:masslambdac} and \ref{tab:massrhoc}. The $\lambda$- and $\rho$-trajectories for the triquarks $((bc)\bar{c})$ and $((bc)\bar{b})$ are shown in Fig. \ref{fig:bc}.

\begin{table}[!phtb]
\caption{Same as Table \ref{tab:masslambda} except for the triquarks $((bc)\bar{c})$ and $((bc)\bar{b})$.}  \label{tab:masslambdac}
\centering
\begin{tabular*}{0.50\textwidth}{@{\extracolsep{\fill}}ccc@{}}
\hline\hline
\addlinespace[2pt]
  $|n^{2s_d+1}l_{j_d},NL\rangle$ & $((bc)\bar{c})$  & $((bc)\bar{b})$ \\
\hline
 $|1^3s_1, 1S\rangle$  &7.97    &11.21      \\
 $|1^3s_1, 2S\rangle$  &8.30    &11.49   \\
 $|1^3s_1, 3S\rangle$  &8.54    &11.68   \\
 $|1^3s_1, 4S\rangle$  &8.74    &11.84   \\
 $|1^3s_1, 5S\rangle$  &8.92    &11.99   \\
 $|1^1s_0, 1S\rangle$  &7.93    &11.17      \\
 $|1^1s_0, 2S\rangle$  &8.26    &11.45   \\
 $|1^1s_0, 3S\rangle$  &8.50    &11.64   \\
 $|1^1s_0, 4S\rangle$  &8.70    &11.80  \\
 $|1^1s_0, 5S\rangle$  &8.88    &11.95   \\
\hline
 $|1^3s_1, 1S\rangle$  &7.98    &11.22   \\
 $|1^3s_1, 1P\rangle$  &8.21    &11.42   \\
 $|1^3s_1, 1D\rangle$  &8.38    &11.56   \\
 $|1^3s_1, 1F\rangle$  &8.53    &11.68   \\
 $|1^3s_1, 1G\rangle$  &8.66    &11.78   \\
 $|1^3s_1, 1H\rangle$  &8.79    &11.88  \\
 $|1^1s_0, 1S\rangle$  &7.94    &11.18   \\
 $|1^1s_0, 1P\rangle$  &8.17    &11.38   \\
 $|1^1s_0, 1D\rangle$  &8.34    &11.52   \\
 $|1^1s_0, 1F\rangle$  &8.49    &11.64  \\
 $|1^1s_0, 1G\rangle$  &8.62    &11.74   \\
 $|1^1s_0, 1H\rangle$  &8.75    &11.84  \\
\hline\hline
\end{tabular*}
\end{table}

\begin{table}[!htbp]
\caption{Same as Table \ref{tab:massrho} except for the triquarks $((bc)\bar{c})$ and $((bc)\bar{b})$.}  \label{tab:massrhoc}
\centering
\begin{tabular*}{0.5\textwidth}{@{\extracolsep{\fill}}ccc@{}}
\hline\hline
\addlinespace[2pt]
  $|n^{2s_d+1}l_{j_d},NL\rangle$ & $((bc)\bar{c})$  & $((bc)\bar{b})$ \\
\hline
 $|1^3s_1, 1S\rangle$  &7.97    &11.21   \\
 $|2^3s_1, 1S\rangle$  &8.31    &11.55   \\
 $|3^3s_1, 1S\rangle$  &8.55    &11.79  \\
 $|4^3s_1, 1S\rangle$  &8.75    &11.99   \\
 $|5^3s_1, 1S\rangle$  &8.94    &12.17  \\
 $|1^1s_0, 1S\rangle$  &7.93    &11.17  \\
 $|2^1s_0, 1S\rangle$  &8.29    &11.53   \\
 $|3^1s_0, 1S\rangle$  &8.53    &11.77   \\
 $|4^1s_0, 1S\rangle$  &8.74    &11.97   \\
 $|5^1s_0, 1S\rangle$  &8.92    &12.16  \\
\hline
 $|1^3s_1, 1S\rangle$  &7.97    &11.21   \\
 $|1^3p_2, 1S\rangle$  &8.23    &11.46   \\
 $|1^3d_3, 1S\rangle$  &8.41    &11.65  \\
 $|1^3f_4, 1S\rangle$  &8.57    &11.81   \\
 $|1^3g_5, 1S\rangle$  &8.71    &11.95   \\
 $|1^3h_6, 1S\rangle$  &8.84    &12.08  \\
 $|1^1s_0, 1S\rangle$  &7.94    &11.18   \\
 $|1^1p_1, 1S\rangle$  &8.21    &11.45  \\
 $|1^1d_2, 1S\rangle$  &8.40    &11.63  \\
 $|1^1f_3, 1S\rangle$  &8.56    &11.79   \\
 $|1^1g_4, 1S\rangle$  &8.70    &11.94   \\
 $|1^1h_5, 1S\rangle$  &8.83    &12.07  \\
\hline\hline
\end{tabular*}
\end{table}

According to Eqs. (\ref{t2q}) and (\ref{pa2qQ}) or Eqs. (\ref{combrt}) and (\ref{pa2qQ}), the Regge slopes decrease with the reduced mass. Consequently, for the $\lambda$-mode excited states, the masses of $((bc)\bar{c})$ are greater than the masses of $((cc)\bar{b})$. While for the $\rho$-mode excited states, the masses of $((bc)\bar{c})$ are smaller than the masses of $((cc)\bar{b})$, see Tables \ref{tab:masslambda}, \ref{tab:massrho}, \ref{tab:masslambdac} and \ref{tab:massrhoc}. For the $\lambda$-mode excited states, the masses of $((bc)\bar{b})$ are smaller than the masses of $((bb)\bar{c})$. While for the $\rho$-mode excited states, the masses of $((bc)\bar{b})$ are greater than the masses of $((bb)\bar{c})$, see Tables \ref{tab:masslambdab}, \ref{tab:massrhob}, \ref{tab:masslambdac} and \ref{tab:massrhoc}.

\begin{table}[!phtb]
\caption{Comparison of theoretical predictions for the spin-averaged masses of the ground state of pentaquarks (in {\gev}).}  \label{tab:masscomp}
\centering
\begin{tabular*}{0.50\textwidth}{@{\extracolsep{\fill}}cccc@{}}
\hline\hline
\addlinespace[2pt]
    & $(\bar{c}(cc))(cc)$    & $(\bar{b}(cc))(cc)$   &  $(\bar{c}(bb))(cc)$ \\
\hline
 Our                     & 7.70   &10.93    &  14.13    \\
 \cite{Rashmi:2024ako}   & 8.55   &11.88    & 15.21     \\
 \cite{Liang:2024met}  & 8.22   &11.46    & 14.62     \\
 \cite{Gordillo:2024blx}  &7.87    & 11.13   &14.30      \\
 \cite{Wang:2021xao}      &7.93    &       &\\
 \cite{Azizi:2024ito}     &7.628   &&\\
 \cite{An:2022fvs}        &8.16  &11.49 &  14.57\\
\hline\hline
\end{tabular*}
\end{table}

\subsection{Discussions}

\begin{figure}[!phtb]
\centering
\includegraphics[width=0.26\textheight]{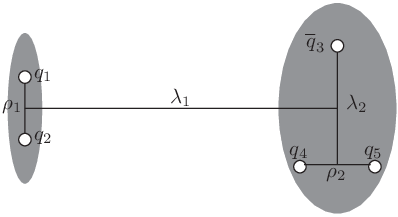}
\caption{Schematic diagram of a pentaquark in the triquark-diquark picture. The left grey part represents a diquark composed of two quarks. The right grey part represents a triquark composed of one antiquark and one diquark. The circles denote quarks and antiquark.}\label{fig:pr}
\end{figure}

Triquarks are colored states and are not observable. The observable is the hadron, for example, the pentaquark. In the triquark-diquark model, one triquark in color $3_c$ and one diquark in color {\cltb} can form a colorless pentaquark. As shown in Fig. \ref{fig:pr}, $\lambda_1$ separates the triquark and the diquark $1$, while $\lambda_2$ separates the antiquark $3$ and the diquark $2$. $\rho_1$ separates quark $1$ and quark $2$ in the diquark $1$, while $\rho_2$ separates quark $4$ and quark $5$ in the diquark $2$. The first two terms in Eq. (\ref{potv}) are determined solely by the nature of the color field; therefore, the same potential can also be used to describe the interaction between diquark and triquark \cite{Lebed:2015tna,Brodsky:2014xia,song2025}. There is $\langle{(\bf{F}_i}\cdot{\bf{F}_j})_{p}\rangle=-4/3$ for the diquark-triquark interaction. By employing the obtained triquark {\rt} relations along with the diquark {\rt} relations, Eqs. (\ref{t2q}) and (\ref{pa2qQ}) or Eqs. (\ref{combrt}) and (\ref{pa2qQ}), we have \cite{song2025}
\begin{align}\label{ppt2q}
M&=m_{R_{\lambda_1}}+\beta_{x_{\lambda_1}}(x_{\lambda_1}+c_{0x_{\lambda_1}})^{2/3}\;(x_{\lambda_1}=L_1,\,N_{r_1}),\nonumber\\
m_{R_{\lambda_1}}&=M_{d_1}+M_{t}+C,\nonumber\\
M_{d_1}&=m_{R_{\rho_1}}+\beta_{x_{\rho_1}}(x_{\rho_1}+c_{0x_{\rho_1}})^{2/3}\;(x_{\rho_1}=l_1,\,n_{r_1}),\nonumber\\
m_{R_{\rho_1}}&=m_{q_1}+m_{q_2}+C/2,\nonumber\\
M_{t}&=m_{R_{\lambda_2}}+\beta_{x_{\lambda_2}}(x_{\lambda_2}+c_{0x_{\lambda_2}})^{2/3}\;(x_{\lambda_2}=L_2,\,N_{r_2}),\nonumber\\
m_{R_{\lambda_2}}&=M_{d_2}+m_{q_3}+C/2,\nonumber\\
M_{d_2}&=m_{R_{\rho_2}}+\beta_{x_{\rho_2}}(x_{\rho_2}+c_{0x_{\rho_2}})^{2/3}\;(x_{\rho_2}=l_2,\,n_{r_2}),\nonumber\\
m_{R_{\rho_2}}&=m_{q_4}+m_{q_5}+C/2,
\end{align}
where
\begin{align}\label{pppa2qQ}
\mu_{\lambda_1}&=\frac{M_{d_1}M_{t}}{M_{d_1}+M_{t}},\;
\mu_{\rho_1}=\frac{m_{q_1}m_{q_2}}{m_{q_1}+m_{q_2}},\nonumber\\
\beta_{L_1}&=\frac{3}{2}\left(\frac{\sigma^2}{\mu_{\lambda_1}}\right)^{1/3}c_{fL_1},\; \beta_{N_{r_1}}=\frac{(3\pi)^{2/3}}{2}\left(\frac{\sigma^2}{\mu_{\lambda_1}}\right)^{1/3}c_{fN_{r_1}},\nonumber\\
\beta_{l_1}&=\frac{3}{2}\left(\frac{\sigma^2}{4\mu_{\rho_1}}\right)^{1/3}c_{fl_1},\; \beta_{n_{r_1}}=\frac{(3\pi)^{2/3}}{2}\left(\frac{\sigma^2}{4\mu_{\rho_1}}\right)^{1/3}c_{fn_{r_1}},\nonumber\\
\mu_{\lambda_2}&=\frac{M_{d_2}m_{q_3}}{M_{d_2}+m_{q_3}},\;
\mu_{\rho_2}=\frac{m_{q_4}m_{q_5}}{m_{q_4}+m_{q_5}},\nonumber\\
\beta_{L_2}&=\frac{3}{2}\left(\frac{\sigma^2}{4\mu_{\lambda_2}}\right)^{1/3}c_{fL_2},\; \beta_{N_{r_2}}=\frac{(3\pi)^{2/3}}{2}\left(\frac{\sigma^2}{4\mu_{\lambda_2}}\right)^{1/3}c_{fN_{r_2}},\nonumber\\
\beta_{l_2}&=\frac{3}{2}\left(\frac{\sigma^2}{4\mu_{\rho_2}}\right)^{1/3}c_{fl_2},\; \beta_{n_{r_2}}=\frac{(3\pi)^{2/3}}{2}\left(\frac{\sigma^2}{4\mu_{\rho_2}}\right)^{1/3}c_{fn_{r_2}}.
\end{align}
$M$, $M_t$, $M_{d_1}$, $M_{d_2}$, $m_{q_1}$, $m_{q_2}$, $m_{q_3}$, $m_{q_4}$, and $m_{q_5}$ are the masses of pentaquark, triquark, diquark $1$, diquark $2$, quark $1$, quark $2$, quark $3$, quark $4$, and quark $5$, respectively. 
Using Eqs. (\ref{ppt2q}) and (\ref{pppa2qQ}) along with (\ref{fitcfxnr}), we crudely estimate the spin-averaged masses of the ground states of pentaquarks $(\bar{c}(cc))(cc)$, $(\bar{b}(cc))(cc)$ and $(\bar{c}(bb))(cc)$. The estimated results are consistent with other theoretical predictions, see Table \ref{tab:masscomp}.

If the triquarks in $3_c$ are too massive, they will fall apart into mesons plus a single quark \cite{Jaffe:2005md}. We argue that the calculation of the masses of highly excited state is useful because it can provide more theoretical data for test by future experimental and theoretical data. And this will be instructive.

According to discussions in \ref{subsec:prelim} and Eqs. (\ref{t2q}) and (\ref{pa2qQ}), for the {\thts}, there are two series of spectra: the spectra of the $\lambda$-excited states and the spectra of the $\rho$-excited states. Correspondingly, there are two series of {\rts}: the $\lambda$-trajectories and the $\rho$-trajectories.
For the {\thts}, the $\lambda$-trajectories and the $\rho$-trajectories have the same behaviors according to Eqs. (\ref{t2q}), (\ref{pa2qQ}), and (\ref{combrt}), $M{\sim}x^{2/3}_{\lambda}$ and $M{\sim}x^{2/3}_{\rho}$, respectively.

In Refs. \cite{Xie:2024lfo,Xie:2024dfe}, it is shown that both the $\lambda$-trajectories and the $\rho$-trajectories for baryons and tetraquarks are concave downwards in the $(M^2,\,x)$ plane. [For the light baryons and tetraquarks, the {\rts} are also concave when the masses of the light constituent are considered.] In this work, we show that both the $\lambda$-trajectories and the $\rho$-trajectories for the {\thts} are also concave downwards in the $(M^2,\,x)$ plane.

To our knowledge, it is the first time that triquarks are discussed using the {\rt} approach. The relation in Eq. (\ref{massf}) \cite{Chen:2022flh}, the potential in Eq. (\ref{potv}) \cite{Ferretti:2019zyh,Bedolla:2019zwg,Ferretti:2011zz,Eichten:1974af,
Chen:2023web,Chen:2023djq}, and the parameter values in Table \ref{tab:parmv} \cite{Feng:2023txx,Faustov:2021qqf} have been used to discuss the {\rts} for mesons, baryons, diquarks, and tetraquarks \cite{Chen:2023djq,Chen:2022flh,Feng:2023txx,Xie:2024lfo,Xie:2024dfe,Song:2025kgw}. They are now employed to study triquarks and pentaquarks, yielding consistent results with other theoretical predictions. This demonstrates the universality of the {\rt} relation and parameter values and illustrates its predictive capability.

\section{Conclusions}\label{sec:conc}
In this work, we attempt to apply the Regge trajectory approach to the triply heavy triquarks $((QQ')\bar{Q}^{\prime\prime})$ $(Q,\,Q',\,Q^{\prime\prime}=b,\,c)$. We present the triquark Regge trajectory relations, and then employ them to crudely estimate the spectra of the {\thts} $((cc)\bar{c})$, $((cc)\bar{b})$, $((bc)\bar{c})$, $((bc)\bar{b})$, $((bb)\bar{c})$, and $((bb)\bar{b})$.

The $\lambda$- and $\rho$-trajectories for the {\thts} are discussed.
The triquark Regge trajectory is a new and very simple approach for estimating the spectra of triquarks. It also provides a simple method to investigate easily the excitations of substructures in pentaquarks and hexaquarks in the triquark picture.

Triquarks are colored states and are not observable. In the diquark-triquark model, the colorless pentaquarks are composed of one triquark and one diquark, and they can be observed. We crudely estimate the spin-averaged masses of the ground states of the pentaquarks $(\bar{c}(cc))(cc)$, $(\bar{b}(cc))(cc)$, and $(\bar{c}(bb))(cc)$ by employing the obtained triquark {\rts}. The estimated results are consistent with other theoretical predictions.

\section*{Acknowledgments}
We are very grateful to the anonymous referees for the valuable comments and suggestions.

\vspace{5mm}
\noindent{\bf Data Availability Statement} This manuscript has no associated data. [Author's comment: All data are included in the manuscript.].

\vspace{5mm}
\noindent{\bf Code Availability Statement} This manuscript has no associated code/software. [Author's comment: Code/Software sharing not applicable to this article as no code/software was generated or analysed during the current study.]

\vspace{5mm}
\noindent{\bf Open Access} This article is licensed under a Creative Commons Attribution 4.0 International License, which permits use, sharing, adaptation, distribution and reproduction in any medium or format, as long as you give appropriate credit to the original author(s) and the source, provide a link to the Creative Commons licence, and indicate if changes were made. The images or other third party material in this article are included in the article's Creative Commons licence, unless indicated otherwise in a credit line to the material. If material is not included in the article's Creative Commons licence and your intended use is not permitted by statutory regulation or exceeds the permitted use, you will need to obtain permission directly from the copyright
holder. To view a copy of this licence, visit \text{http://creativecommons.org/licenses/by/4.0/}.

\noindent{Funded by SCOAP$^3$.}  

\appendix

\section{States of triquarks}
The states of triquarks in the diquark picture are listed in Table \ref{tab:bqstates}.

\begin{table*}[!phtb]
\caption{The states of triquarks in $3_c$ composed of one diquark in $\cltba$ and one antiquark in $\cltba$. The notation is explained in \ref{subsec:prelim}. Here, $q$, $\qp$ and $\qpp$ represent both the light quarks and the heavy quarks.}  \label{tab:bqstates}
\centering
\begin{tabular*}{\textwidth}{@{\extracolsep{\fill}}ccccc@{}}
\hline\hline
$J^P$ & $(L,l)$  &  Configuration \\
\hline
$\frac{1}{2}^+$ & $(0,0)$ & $\left({\qqs}^{\cltba}_{n^1s_0}{\qpp}\right)^{3_c}_{N^2S_{1/2}}$,\;
$\left({\qqb}^{\cltba}_{n^3s_1}{\qpp}\right)^{3_c}_{N^2S_{1/2}}$,\\
& $(1,1)$ &
$\left({\qqb}^{\cltba}_{n^1p_1}{\qpp}\right)^{3_c}_{N^2P_{1/2}}$,\;
$\left({\qqb}^{\cltba}_{n^1p_1}{\qpp}\right)^{3_c}_{N^4P_{1/2}}$,\;
$\left({\qqs}^{\cltba}_{n^3p_0}{\qpp}\right)^{3_c}_{N^2P_{1/2}}$,\;
$\left({\qqs}^{\cltba}_{n^3p_1}{\qpp}\right)^{3_c}_{N^2P_{1/2}}$,\\
& &
$\left({\qqs}^{\cltba}_{n^3p_1}{\qpp}\right)^{3_c}_{N^4P_{1/2}}$,\;
$\left({\qqs}^{\cltba}_{n^3p_2}{\qpp}\right)^{3_c}_{N^4P_{1/2}}$\\
&$\cdots$ &$\cdots$ \\
$\frac{1}{2}^-$ & $(1,0)$ &
$\left({\qqs}^{\cltba}_{n^1s_0}{\qpp}\right)^{3_c}_{N^2P_{1/2}}$,\;
$\left({\qqb}^{\cltba}_{n^3s_1}{\qpp}\right)^{3_c}_{N^2P_{1/2}}$,\;
$\left({\qqb}^{\cltba}_{n^3s_1}{\qpp}\right)^{3_c}_{N^4P_{1/2}}$,\\
& $(0,1)$ &
$\left({\qqb}^{\cltba}_{n^1p_1}{\qpp}\right)^{3_c}_{N^2S_{1/2}}$,\;
$\left({\qqs}^{\cltba}_{n^3p_0}{\qpp}\right)^{3_c}_{N^2S_{1/2}}$,\;
$\left({\qqs}^{\cltba}_{n^3p_1}{\qpp}\right)^{3_c}_{N^2S_{1/2}}$,\\
&$\cdots$ &$\cdots$ \\
$\frac{3}{2}^+$ & $(0,0)$ &
$\left({\qqb}^{\cltba}_{n^3s_1}{\qpp}\right)^{3_c}_{N^4S_{3/2}}$,\\
& $(1,1)$ &
$\left({\qqb}^{\cltba}_{n^1p_1}{\qpp}\right)^{3_c}_{N^2P_{3/2}}$,\;
$\left({\qqb}^{\cltba}_{n^1p_1}{\qpp}\right)^{3_c}_{N^4P_{3/2}}$,\;
$\left({\qqs}^{\cltba}_{n^3p_0}{\qpp}\right)^{3_c}_{N^2P_{3/2}}$,\;
$\left({\qqs}^{\cltba}_{n^3p_1}{\qpp}\right)^{3_c}_{N^2P_{3/2}}$,\;\\
& &
$\left({\qqs}^{\cltba}_{n^3p_1}{\qpp}\right)^{3_c}_{N^4P_{3/2}}$,\;
$\left({\qqs}^{\cltba}_{n^3p_2}{\qpp}\right)^{3_c}_{N^4P_{3/2}}$,\;
$\left({\qqs}^{\cltba}_{n^3p_2}{\qpp}\right)^{3_c}_{N^6P_{3/2}}$,\\
&$\cdots$ &$\cdots$ \\
$\frac{3}{2}^-$ & $(1,0)$ &
$\left({\qqs}^{\cltba}_{n^1s_0}{\qpp}\right)^{3_c}_{N^2P_{3/2}}$,\;
$\left({\qqb}^{\cltba}_{n^3s_1}{\qpp}\right)^{3_c}_{N^2P_{3/2}}$,\;
$\left({\qqb}^{\cltba}_{n^3s_1}{\qpp}\right)^{3_c}_{N^4P_{3/2}}$,\\
& $(0,1)$ &
$\left({\qqb}^{\cltba}_{n^1p_1}{\qpp}\right)^{3_c}_{N^4S_{3/2}}$,\;
$\left({\qqs}^{\cltba}_{n^3p_1}{\qpp}\right)^{3_c}_{N^4S_{3/2}}$,\;
$\left({\qqs}^{\cltba}_{n^3p_2}{\qpp}\right)^{3_c}_{N^4S_{3/2}}$\\
&$\cdots$ &$\cdots$ \\
\hline\hline
\end{tabular*}
\end{table*}

\end{document}